\newcommand{\bee}{\begin{equation}}
\newcommand{\ene}{\end{equation}}
\newcommand{\beea}{\begin{eqnarray}}
\newcommand{\enea}{\end{eqnarray}}

\documentclass[aps,preprint,pre,showpacs,superscriptaddress]{revtex4}
\usepackage{amsmath} 
\usepackage{rotating}
\usepackage{graphicx}
\usepackage{dcolumn}
\usepackage{bm}
\begin{document}
\title{Ground State Energy of Hydrogen-Like Ions in Quatum Plasmas}

\author{M. Akbari-Moghanjoughi}
\address{Faculty of Sciences, Department of Physics, Azarbaijan Shahid Madani University, 53714-161 Tabriz, Iran}
\author{Alireza Abdikian}
\address{Department of Physics, Malayer University, Malayer 65719-95863, Iran}
\author{Arash Phirouznia}
\address{Faculty of Sciences, Department of Physics, Azarbaijan Shahid Madani University, 53714-161 Tabriz, Iran}
\address{Condensed Matter Computational Research Lab., Azarbaijan Shahid Madani University, 53714-161 Tabriz, Iran}

\begin{abstract}
Using the asymptotic iteration method (AIM) we investigate the variation in the 1s energy levels of hydrogen and helium-like static ions in fully degenerate electron gas. The semiclassical Thomas-Fermi (TF), Shukla-Eliasson (SE) and corrected Shukla-Eliasson (cSE) models are compared. It is remarked that these models merge into the vacuum level for hydrogen and helium-like ions in the dilute classical electron gas regime. While in the TF model hydrogen ground state level lifts monotonically towards the continuum limit with increase in the electron concentration, in the SE and cSE models universal bound stabilization valley through the energy minimization occurs at a particular electron concentration range for the hydrogen-like ion which for cSE model closely matches the electron concentrations in typical metals. The later stabilizing mechanism appears to be due to the interaction between plasmon excitations and the Fermi lengthscales in metallic density regime. In the case of helium-like ions, however, no such stability mechanism is found. The application of cSE model with electron exchange and correlation effects reveals that cSE model qualitatively accounts for the number-density and lattice parameters of elemental metals within the framework of free electron assumption. According to the cSE model of static charge screening a simple metal-insulator transition criterion is defined. Current investigation may further elucidate the underlying physical mechanisms in the formation and dielectric properties of metallic compounds.
\end{abstract}

\date{\today}

\maketitle
\section{Introduction}

Electrostatic and electrodynamic response of electron gas to external perturbations is one of principal subjects in many areas of basic scientific research such as in plasmas, solid state physics, optics, nanotechnology, plasmonics, low dimensional systems, etc. \cite{lucio,mark,haug,gardner,man1,maier,yofee}. The dynamic structure factor of a statistical ensemble, as a fundamental element of the linear response theory, provides broad information on dynamic density-density correlations, inelastic scattering of different types, resonant absorbtion, dynamic charge screening and ion stopping power of the particle system via the fluctuation-dissipation theorem \cite{ichimaru1,ichimaru2,ichimaru3,sturm}. The dielectric response of bound electrons has fundamental impact on almost all physical properties of crystalline and amorphous solids from the optical to electric and thermal aspects \cite{kit,ash}. In metallic compounds and semiconductors \cite{hu1,seeg}, however, the nearly free electron gas is responsible for band gap formation and many outstanding characteristic electrical, optical and magnetic properties due to quantum statistical effects. On the other hand, the static structure factor gives vital information on the ion-ion correlation strength, elastic scattering of electromagnetic radiation from crystalline material, and pair distribution function which provides useful information on liquid-solid phase transitions \cite{ionst}. The later is also closely related to the static screening potential around the test charge \cite{zhao,mold0,hong,ydj1,ydj2,yoon}.

The most basic theory of dielectric function and quantum static charge screening in the free electron gas is due to the Thomas-Fermi (TF) theory. This theory uses the Fermi distribution function in order to fully account for the quantum statistical effect. However, the TF screening theory is an ikonic representation of Debye-like exponential screening effect which is based on the single electron wavefunction. Due to neglect of the collective effects which is caused by single electron interaction with a self consistent (Hartree) electrostatic field, as is considered in the Lindhard's response theory of random phase approximation (RPA) \cite{lind,stern}, the TF theory does not account for important many-body effects \cite{fetter,mahan,pin} such as Friedel oscillations. However, the gradient corrected version of TF theory has been shown to capture many essential features of the quantum dielectric response and charge screening \cite{sm,michta}. On the other hand, recent developments in quantum kinetic and hydrodynamic (HD) \cite{haas1,manfredi,hurst,man3} models has provided alternative method of quantum dielectric response measure. Study of static charge screening using linearized HD model leads to overestimated account for the quantum Bohm potential \cite{mold,akbhd,hasmah}. However, the kinetic corrected quantum potential term in HD model has been shown to lead to identical results to the gradient corrected TF for the quantum static charge screening \cite{elak}.

For a classical gas with increase in the temperature or number density of species the increase in the collision frequency between different species can lead to the ionization and plasma formation \cite{chen,krall}. The Saha criterion provides the estimate of such ionization in given thermodynamic equilibrium \cite{saha}. In the case of quantum plasmas where the interparticle spacing compares to the de Broglie thermal wavelength the increase in the electron number density is the dominant cause of pressure ionization in degenerate matter \cite{bonitz}. However, the electron-electron collisions are prohibited by the Pauli exclusion mechanism in dense quantum plasmas \cite{man2} and can not contribute to the ionization. To investigate the ionization problem in quantum plasmas it is a good practice to study the atomic energy levels in the quantum electron gas in order to see the effect of electron density on the variation of bound state with the increase in the density. There has been an increased attention to this subject over the past few years \cite{yu,sahoo,kar,paul1,paul2,paul3,paul4,paul5,soy1,soy2}. Use of different numerical algorithms such as the Rayleigh–Ritz variational approach \cite{mac} and asymptotic iteration method (AIM) \cite{cif1,cif2} confirms that the energy of atomic levels when placed in extreme condition such as high density or temperature are shifted towards the continuum limit. It is seen that more and more bound states are lost as the number density of electrons increases until all energy levels become unbound. Recent investigation on quantum plasmas \cite{hu} with five different screening potentials reveals a quite ramarkable behavior for newly suggested potentials by Shukla and Eliasson (SE) and kinetic corrected Shukla-Eliasson (cSE) version. It is particularly shown that for the cSE potential around the metallic density of electrons the 1s energy level of hydrogen and helium move away from the continuum and become more bound. These finding, if it is further confirmed, can suggest novel stabilization and binding mechanism for the metallic electron concentration regime. It is however the aim of current study in order to investigate the atomic bound states for a much wider electron concentration in order to see if or not such major deviations from standard screening scheme is present for these new models.

\section{Static Charge Screening}

The most familiar and simplest static charge screening effect corresponds to the classical Debye shielding model which gives rise to the potential around the screened charge $Q=Ze$ ($Z$ being the ionization number) as $\Phi_D=Q\exp(-r/\lambda_D)/r$ where $\lambda_D=\sqrt{k_B T_e/(4\pi e^2 n_0)}$ with $T_e$ and $n_0$ being the screening electron fluid temperature and number density, respectively. This theory relies on the Maxwell-Boltzmann distribution function which predicts an exponential energy-density relation $n(r)=n_0(r)\exp(e\Phi/k_B T)$ in a thermal equilibrium. However, as the electron gas around the impurity charge increases to the extent of critical length-scale $\Lambda_D\simeq n_0^{-1/3}$ in which $\Lambda_D=h/\sqrt{2\pi m_e k_B T_e}$ is the de Broglie thermal wavelength, the overlap of single-electron wavefunctions lead to quantum effects deviating the statistical description of the system from standard Maxwell-Boltzmann theory. The quantum regime starts at approximate density of $n_0\simeq 10^{18}$cm$^{-3}$ with the criteria of quantum coupling parameter being the ratio of the potential-to-kinetic energy ration approaching unity, i.e., $Q_c\simeq 1$. Other criteria may be given based on the Landau length $\Lambda_L=e^2/k_B T_e$ where $\lambda_L\ge d$ ($d\simeq n_0^{-1/3}$ being the average inter-particle distance) or equivalently $k_B T_e\le e^2/d$ coins the quantum realm. Therefore, the criteria of application of quantum models to a plasma at equilibrium state may be either $d\le \Lambda_L$ or $d\le \Lambda_D$. The simplest quantum static screening model for plasmas is the well-known Thomas-Fermi model which relies on the generalized energy density relation $n_e(r)=n_0(r){\rm{L}}{{\rm{i}}_{3/2}}\left\{ { - \exp \left[ {\left( {\mu  - e\Phi } \right)/{k_B}{T_e}} \right]} \right\}/{\rm{L}}{{\rm{i}}_{3/2}}\left[ { - \exp \left( {\mu /{k_B}{T_e}} \right)} \right]$ where the polylog function $\text{Li}$ is defined through the Fermi integrals as
\begin{equation}\label{li}
{\rm{Li}}_{\nu}( - {{\rm{e}}^z}) = -\frac{1}{\Gamma (\nu)}\mathop \smallint \limits_0^\infty  \frac{{{x^{\nu-1}}}}{{\exp (x - z) + 1}}{\rm{d}}x,\hspace{3mm}\nu > 0,
\end{equation}
where $\Gamma$ is the ordinary gamma function and $\mu$ stands for the chemical potential of the electron gas. Note that in the fully degenerate electron gas limit, $z\gg 1$, we have $\lim_{z\rightarrow\infty} {\rm Li}_{\nu}(-e^z)=-z^\nu/\Gamma(\nu+1)$ and in the classical limit, $z \ll -1$, we have ${\rm Li_\nu}(-e^z)\approx-e^z$. The full degeneracy starts when $T_e\ll T_{F}$ in which $T_F=E_E/k_B$ is the Fermi temperature with $E_F=\hbar^2(3\pi^2 n_0)^{2/3}/2m_e$ being the Fermi energy. In the fully degenerate limit the statistical description of quantum electron gas becomes completely independent of the electron fluid temperature and it only depends on the number-density of electrons in the Fermi gas. The screening potential of Thomas-Fermi model is essentially similar to the Debye model $\Phi_{TF}=Q\exp(-r/\lambda_{TF})/r$ with the new definition of the Thomas-Fermi length as $\lambda_{TF}=1/\sqrt{4\pi e^2\partial{n_e}/\partial{\mu}}$ in which the number density is given by \cite{elak}
\begin{equation}\label{iso}
{n_e} =  - N{\rm{L}}{{\rm{i}}_{3/2}}\left[ { - {\rm{exp}}\left( {\mu/k_B T_e} \right)} \right].
\end{equation}
where $N = {2}/{{\Lambda_D^3}}$. The Thomas-Fermi wavenumber for arbitrary degenerate electron gas reads $k_{TF}=1/\lambda_{TF}$. In the full degeneracy limit the effect of temperature in (\ref{iso}) becomes insignificant and the chemical potential equals the Fermi-energy and therefore one obtains the Thomas-Fermi wavenumber $k_{TF}=\sqrt{4 m_e e^2k_F/\pi^2\hbar^2}$ in which $k_F=\sqrt{2m_eE_F}/\hbar$ is the Fermi wavenumber. A simple expression for the Thomas-Fermi wavenumber at zero-temperature limit is $k_{TF}=\sqrt{3}\omega_p/v_F$ where $\omega_p=\sqrt{4\pi e^2 n_0/m_e}$ and $v_F=\sqrt{2E_F/m_e}$ are plasmon frequency and electron Fermi speed, respectively.

The Thomas-Fermi screening theory is a semi-classical one which appropriately incorporates the quantum statistical effects to some extent. For instance, the quantum recoil or diffraction effect due to the quantum potential is ignored in this model. The quantum potential also known as the Bohm potential is understood to be the origin of many nonlocal effect in dielectric response of degenerate electron gas. One appropriate alternative to the conventional Thomas-Fermi models is the quantum kinetic theory which is based on the Wigner-transformation. The quantum hydrodynamic \cite{haas1} which is obtained from the moments of the Wigner-Poisson system \cite{man2} is the simplest yet a powerful variation among other theories in order to investigate plasma response through the interactions of electron via consistent electromagnetic potentials defined by the Maxwell equations. The gradient corrected Thomas-Fermi model is also another equivalent alternative to the original Thomas-Fermi model which correctly accounts for the quantum electron diffraction effect. Using the quantum hydrodynamic model Shukla and Eliasson \cite{seprl} for the first time calculated the dielectric function and the static charge screening potential in a completely degenerate electron gas
\begin{equation}\label{scr}
\Phi (r) = \frac{Q}{{2{\pi ^2}}}\int_{\bf k} {\frac{{\exp (i{\bf{k}}\cdot{\bf{r}}){\bf{dk}}}}{{{k^2}\varepsilon (k,0)}}},
\end{equation}
where $\varepsilon (k,0)$ is the static longitudinal dielectric response of electron gas and
\begin{equation}\label{hdd}
\varepsilon (k,0) = 1 + \frac{{\omega _p^2}}{{{k^2}v_F^2/3 + {\hbar ^2}{k^4}/4m_e^2}},
\end{equation}
where the electron exchange and correlation and effective mass contributions has been ignored for simplicity by setting $v_{XC}=0$ and $m^*=m_e$. They obtained the simple analytic expression for screening potential of cosine-sine-exponential form as follows
\begin{equation}\label{se}
{\Phi _{SE}}(r) = \frac{{Q\exp ( - {A_{SE}}r)}}{r}\left[ {\cos ({B_{SE}}r) + {b_{SE}}\sin ({B_{SE}}r)} \right],
\end{equation}
where the potential parameters are given as
\begin{equation}\label{separ}
{A_{SE}} = {k_{TF}}\frac{{\sqrt {\sqrt {4{\alpha _{SE}}} +1} }}{{\sqrt {4{\alpha _{SE}}} }},\hspace{3mm}{B_{SE}} = {k_{TF}}\frac{{\sqrt {\sqrt {4{\alpha _{SE}}}-1 } }}{{\sqrt {4{\alpha _{SE}}} }},\hspace{3mm}{b_{SE}} = \frac{1}{{\sqrt {4{\alpha _{SE}} - 1} }}.
\end{equation}
The parameter of the potential $\alpha_{SE}=3\hbar^2\omega_p^2/(4m_e^2 v_F^4)$ is critical to the shape of the screening potential which can be either monotonic when $\alpha_{SE}<1/4$ or oscillatory when $\alpha_{SE}>1/4$. Note also that the single potential (\ref{se}) with the given parameters in (\ref{separ}) is valid for both potential forms of monotonic and oscillatory. Because of discrepancy between the above result and the density functional theory (DFT) simulations an intense debate has gone on the validity of the hydrodynamic or DFT theories \cite{seprl,bonitz1,sea1,bonitz2,sea2,bonitz3}. This is because the SE potential gives rise to the Lennard-Jones type attractive potential around the screened ion for a wide range of the electron number density which is well above the metallic density regime. The latter phenomenon leads to significant consequences for the inertial confinement scheme at superdense plasma regime and warm dense matter (WDM) \cite{ko}.

\begin{sidewaysfigure}[ptb]\label{Figure1}
\includegraphics[scale=0.8]{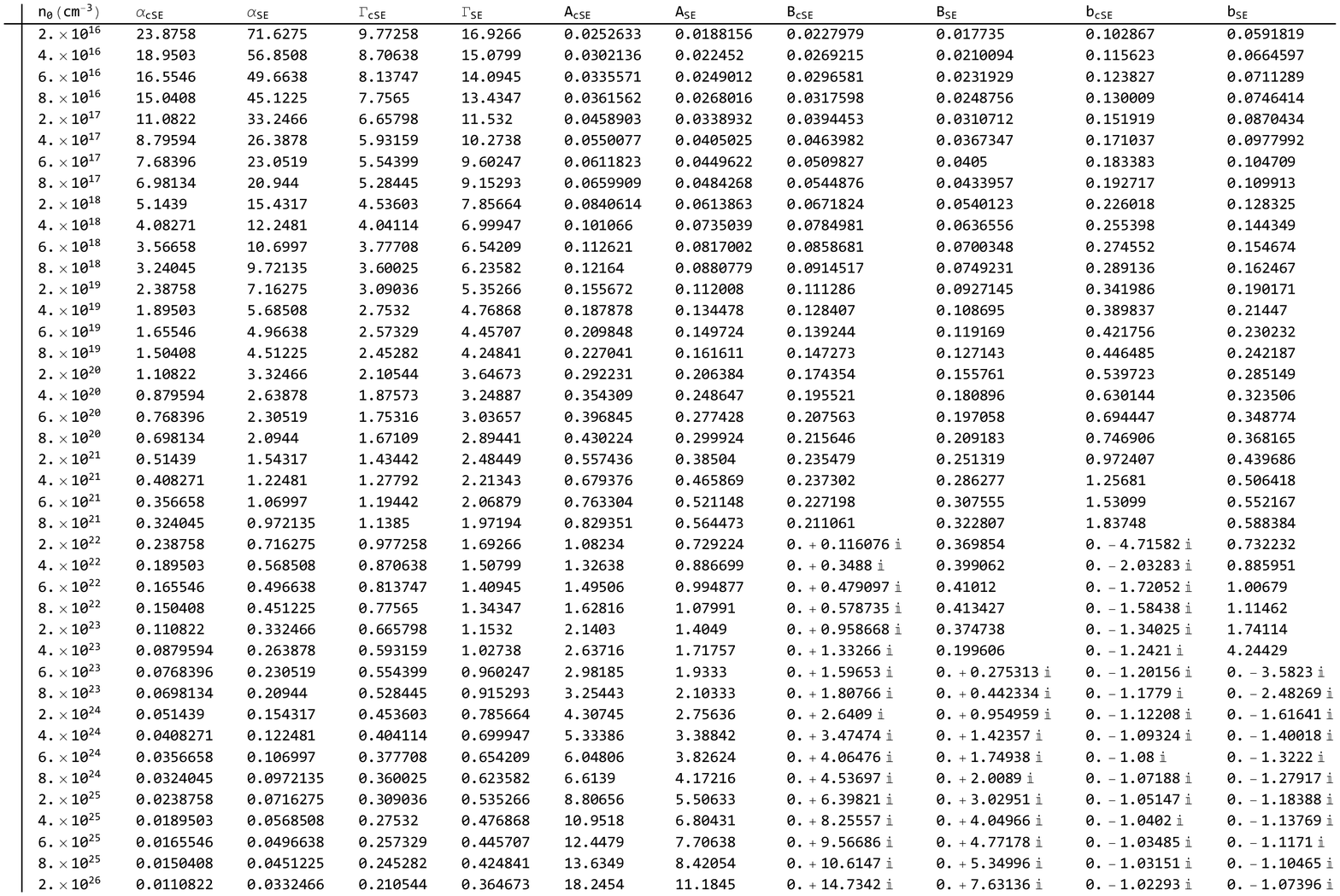}\caption{Different parameters for the screening potential in the SE and cSE models for a wide range of electron number density values.}
\end{sidewaysfigure}

On the other hand, using the Lindhard theory of dielectric response based on RPA and quantum kinetic theory in the zero temperature electron gas limit the dielectric function with corrected $\alpha$-parameter for hydrodynamic formulation has been recently obtained which leads to the correct form of screening potential in the low phase speed limit of excitations \cite{akbhd}
\begin{equation}\label{am}
{\Phi _{cSE}}(r) = \frac{{Q\exp ( - {A_{cSE}}r)}}{r}\left[ {\cos ({B_{cSE}}r) + {b_{cSE}}\sin ({B_{cSE}}r)} \right],
\end{equation}
with the potential parameters given as
\begin{equation}\label{ampar}
{A_{cSE}} = {k_{TF}}\frac{{\sqrt {\sqrt {4{\alpha _{cSE}}} +1} }}{{\sqrt {4{\alpha _{cSE}}} }},\hspace{3mm}{B_{cSE}} = {k_{TF}}\frac{{\sqrt {\sqrt {4{\alpha _{cSE}}} -1} }}{{\sqrt {4{\alpha _{cSE}}} }},\hspace{3mm}{b_{cSE}} = \frac{1}{{\sqrt {4{\alpha _{cSE}} - 1} }}.
\end{equation}
The correction applies only to the parameter, $\alpha$, leading to the expression $\alpha_{cSE}=\hbar^2\omega_p^2/(4m_e^2 v_F^4)$.  However, the mathematical structure of screening potentials (\ref{se}) and (\ref{am}) may be further simplified as
\begin{equation}\label{gamma}
\Phi (r) = \frac{{Z\exp ( - {A_\Gamma }r)}}{r}\left[ {\cos ({B_\Gamma }r) + {b_\Gamma }\sin ({B_\Gamma }r)} \right].
\end{equation}
with the potential parameters given as
\begin{equation}\label{gammapar}
{A_\Gamma } = \frac{{{k_{TF}}}}{{{k_0}}}\frac{{\sqrt {\Gamma +1} }}{\Gamma },\hspace{3mm}{B_\Gamma } = \frac{{{k_{TF}}}}{{{k_0}}}\frac{{\sqrt {\Gamma -1} }}{\Gamma },\hspace{3mm}{b_\Gamma } = \frac{1}{{\sqrt {{\Gamma ^2} - 1} }}.
\end{equation}
where $\Gamma_{cSE}=E_p/2E_F$ with $E_p=\hbar\omega_p$ being the plasmon energy and $\Gamma_{SE}=\sqrt{3}\Gamma_{cSE}$. Note that the potential (\ref{gamma}) is normalized in Rydberg energy unit with $r$ being scaled to the Bohr radius $r_B=\hbar^2/m_e e^2$ and $k_0=1/r_B$. The critical screening ($\Gamma=1$) in cSE-model \cite{mold0}, corresponds to the point where plasmon energy becomes twice the Fermi energy, or equivalently, when the plasmon wavenumber equals $\sqrt{2}$ times the Fermi-wavenumber. This is the point where the screening potential turns from monotonic to oscillatory and viceversa. The critical screening in the cSE model corresponds to the electron number density of $n_0\simeq 64/(81\pi^5 r_B^3)=1.74\times 10^{22}$cm$^{-3}$ beyond which the potential becomes oscillatory. In the SE-model the critical point coincides with the electron number density of $n_0\simeq4.7\times 10^{23}$cm$^{-3}$ which is almost one order of magnitude larger. The potential parameters of SE and cSE model are compared in Table 1 (Fig. 1) for a wide range of electron number density.

\begin{figure}[ptb]\label{Figure2}
\includegraphics[scale=0.6]{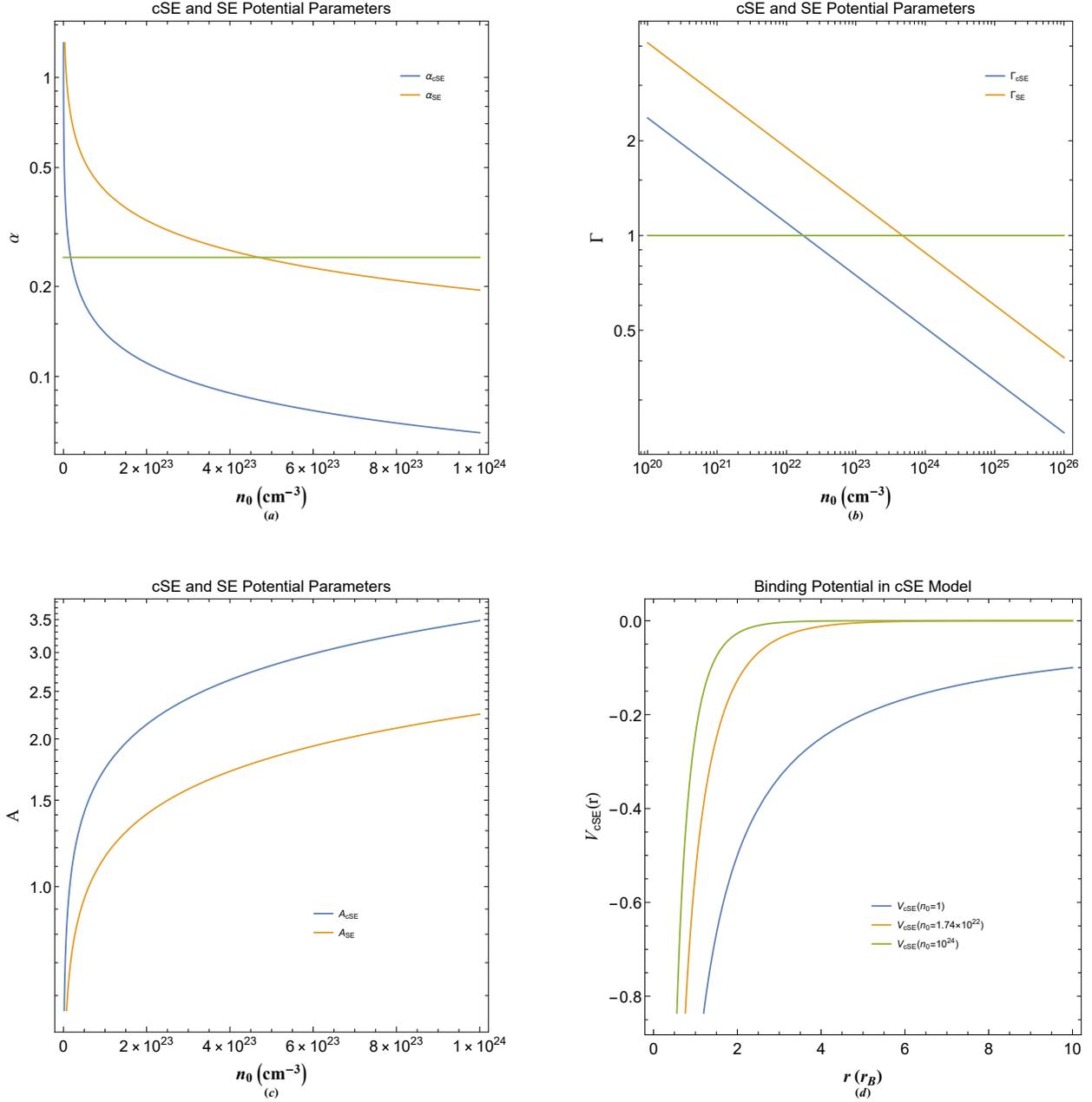}\caption{Variations of the potential parameters for charge screening in SE and cSE models. The horizontal line in plots (a) and (b) indicate the critical screening values for potential parameters.}
\end{figure}

Figure 2 shows the variation of potential parameters for the SE and cSE screening models. Figure 2(a) depicts the variations in $\alpha$ parameters. The critical point is indicated by a horizontal line at $\alpha=1/4$. It is seen that for the whole range of electron density the corresponding parameter for SE model is larger than that of the cSE and intersects the critical value at larger electron number density. Moreover, Fig. 2(b) shows the parameter $\Gamma$ for these screening models in a logarithmic scale. It is clearly remarked that the gamma parameter meets the critical value for SE model at larger electron density than for the cSE model. Figure 2(c) shows the variations in the screening wavenumbers $A$ in both models. It is clear that the screening length of the cSE model is quite smaller compared to the SE model for the whole range of electron number density. For both models it decreases with increase of the number density, as expected. The bound potential of the cSE model is shown in Fig. 2(d) for over-critical, critical and under-critical electron density values. The lowest potential corresponds to the isolated hydrogen bound potential. The 1s bound state of the potential is $-13.6$eV. However as the hydrogen-like ion is introduced in a dense electron fluid the bound potential shrinks repelling the bound levels towards the continuum limit.

The effect of exchange and correlations which is electron-electron interactions are not included in the standard RPA dielectric function of Lindhard which has been used to obtain the cSE quantum screening potential. However, in order to include these important effects on the dielectric response which play fundamental role on the physical properties of metals we follow the linearized hydrodynamics treated in Ref. \cite{seprl}. The exchange-correlation potential may be written as
\begin{equation}\label{xcp}
{V_{XC}}({n_0}) = 0.985{e^2}n_0^{1/3}\left[ {1 + \frac{{0.034}}{{{r_B}n_0^{1/3}}}\ln \left( {1 + 18.37{r_B}n_0^{1/3}} \right)} \right].
\end{equation}
For the cSE model which includes the electron exchange and correlation effects we find the potential parameters as
\begin{equation}\label{xcpar}
{A_{XC}} = \frac{{{k_{TF}}}}{{{k_0}}}\frac{{\sqrt {{\Gamma _{XC}} + 1} }}{{{\Gamma _{XC}}}},\hspace{3mm}{B_{XC}} = \frac{{{k_{TF}}}}{{{k_0}}}\frac{{\sqrt {{\Gamma _{XC}} - 1} }}{{{\Gamma _{XC}}}},\hspace{3mm}{b_{XC}} = \frac{1}{{\sqrt {\Gamma _{XC}^2 - 1} }},
\end{equation}
where the generalized screening parameter in the cSE model including exchange-correlation effects reads
\begin{equation}\label{xcgamma}
{\Gamma _{XC}} = \frac{{{E_p}}}{{2({E_F} + {E_{XC}})}},\hspace{3mm}{E_{XC}} = \frac{1}{2}m_e v_{XC}^2,\hspace{3mm}{v_{XC}} = \sqrt {0.328\frac{{{e^2}{n_0^{1/3}}}}{{2{m_e}}}\left( {1 + \frac{{0.62}}{{1 + 18.36{r_B}{n_0^{1/3}}}}} \right)}.
\end{equation}

\section{Energy Eigenvalues For TF, SE and cSE Models}

\begin{figure}[ptb]\label{Figure3}
\includegraphics[scale=0.8]{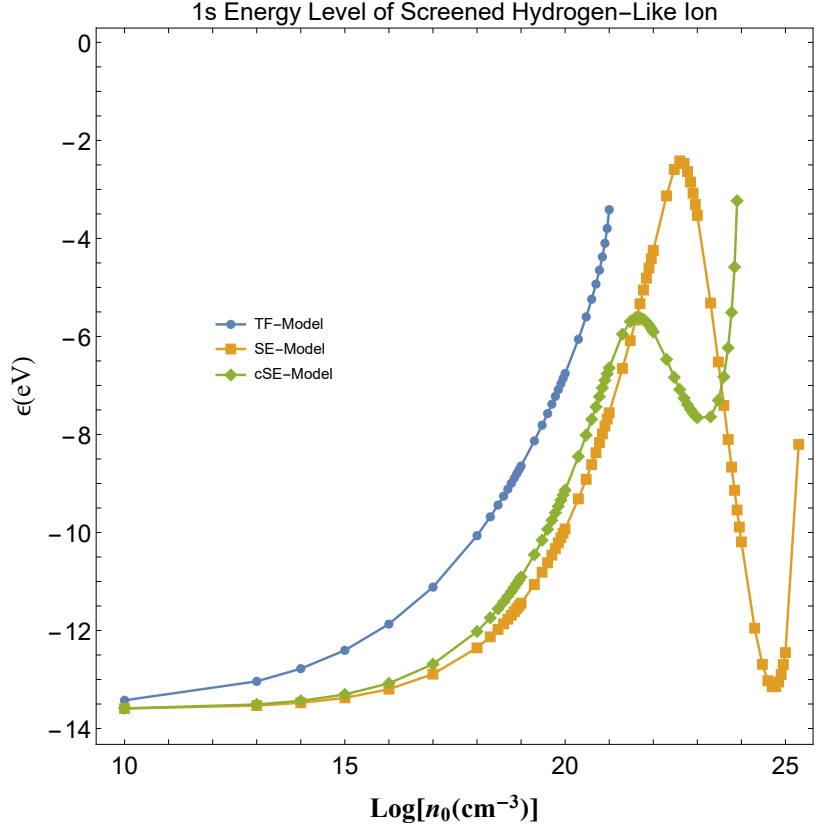}\caption{The variation of 1s energy level of hydrogen-like ion immersed in an electron fluid with different electron number density in a logarithmic scale in three different screening models, namely, the TF, SE, and cSE models.}
\end{figure}

\begin{figure}[ptb]\label{Figure4}
\includegraphics[scale=0.8]{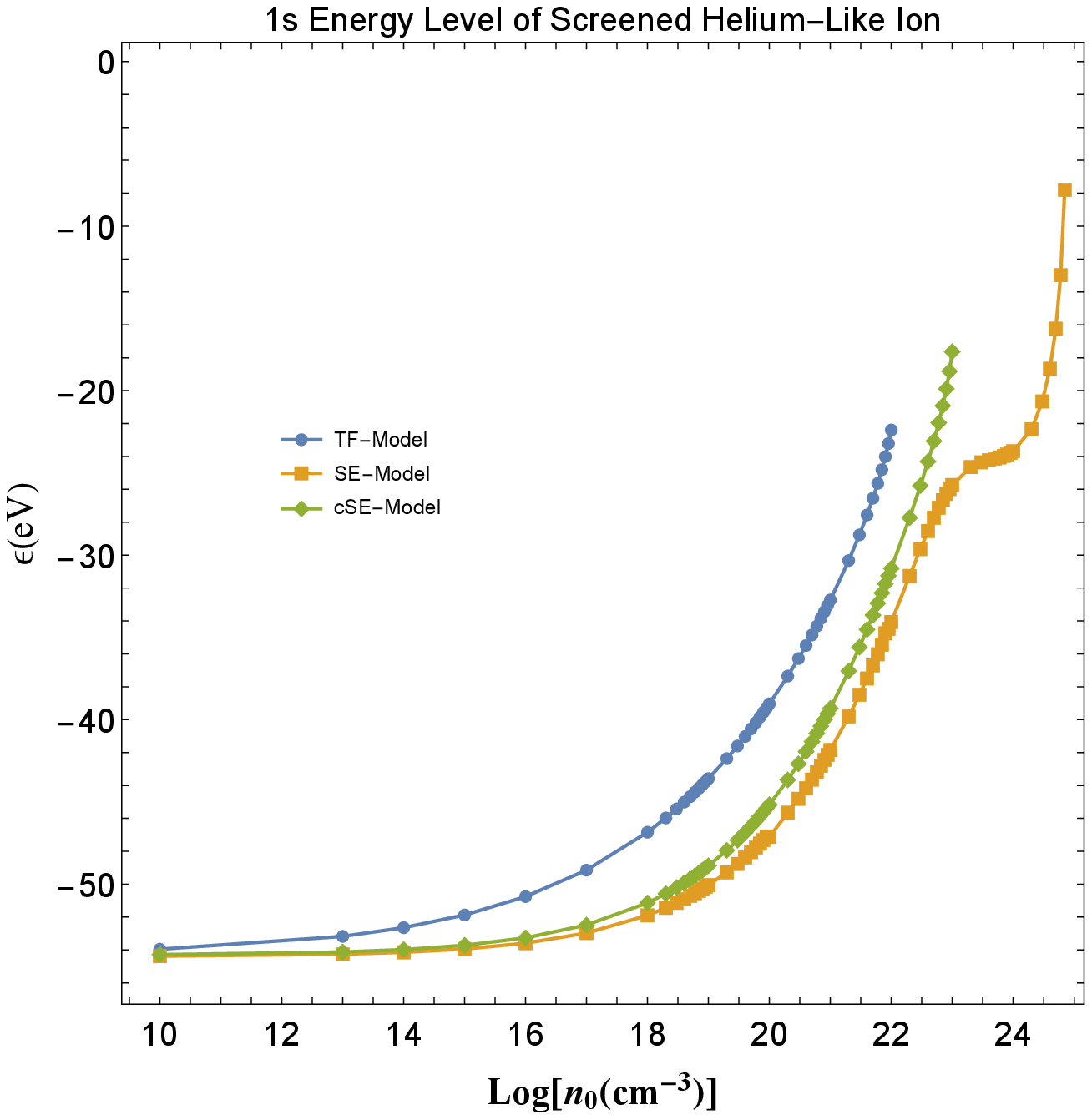}\caption{The variation of 1s energy level of helium-like ion immersed in an electron fluid with different electron number density in a logarithmic scale in three different screening models, namely, the TF, SE and cSE model.}
\end{figure}

\begin{sidewaysfigure}[ptb]\label{Figure5}
\includegraphics[scale=0.8]{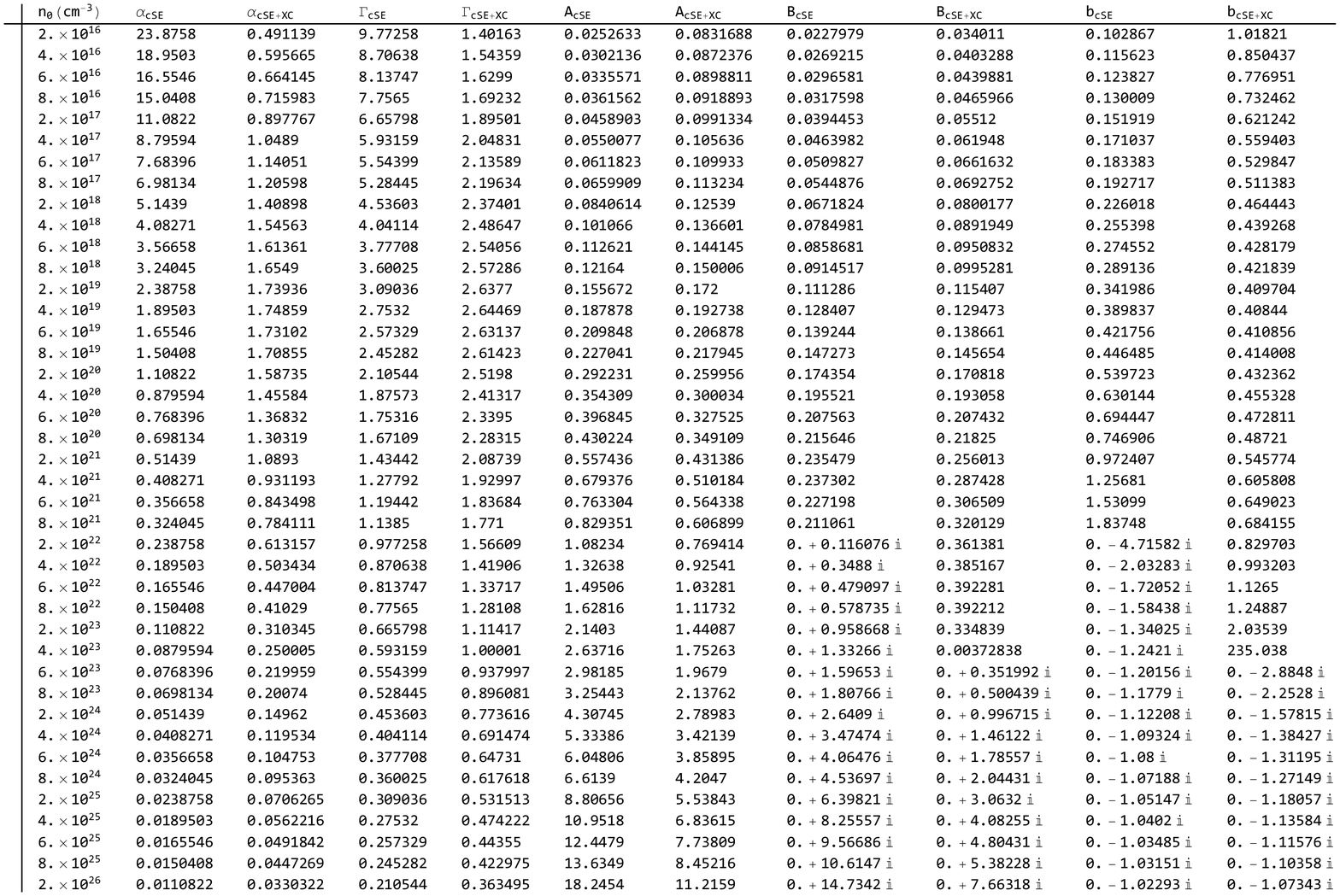}\caption{Different parameters for the screening potential in the cSE models with and without exchange effect for a wide range of electron number density values.}
\end{sidewaysfigure}

\begin{figure}[ptb]\label{Figure6}
\includegraphics[scale=0.6]{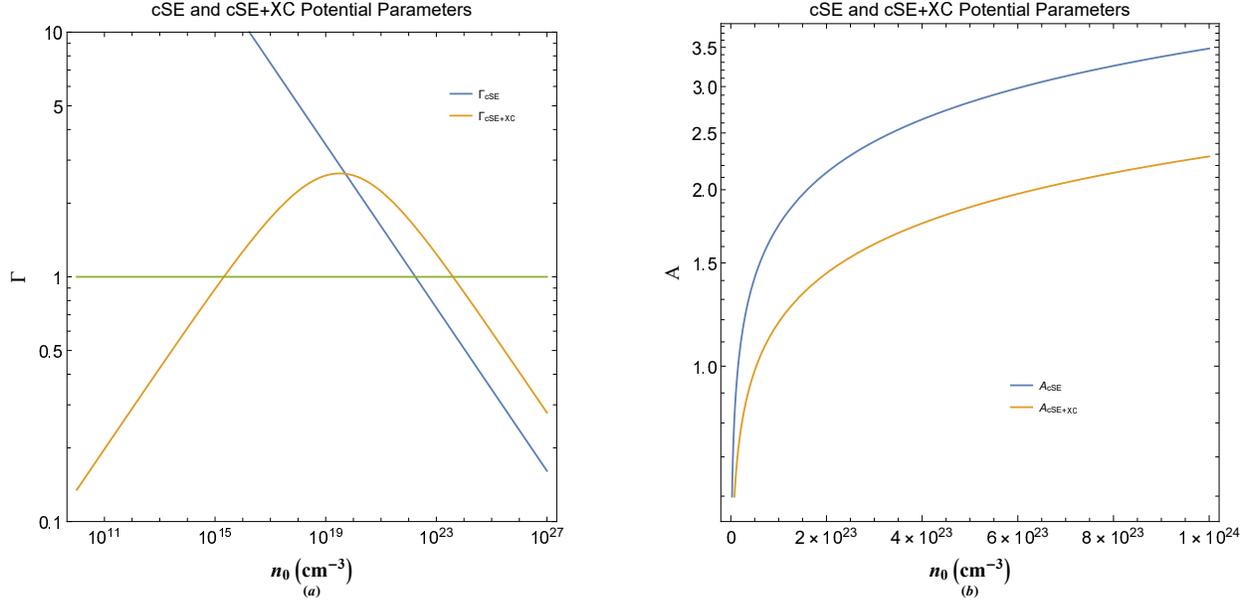}\caption{Variations of the potential parameters for charge screening in the cSE model with and without exchange effect. The horizontal line in plots (a) indicate the critical screening values for potential parameters.}
\end{figure}

\begin{figure}[ptb]\label{Figure7}
\includegraphics[scale=0.8]{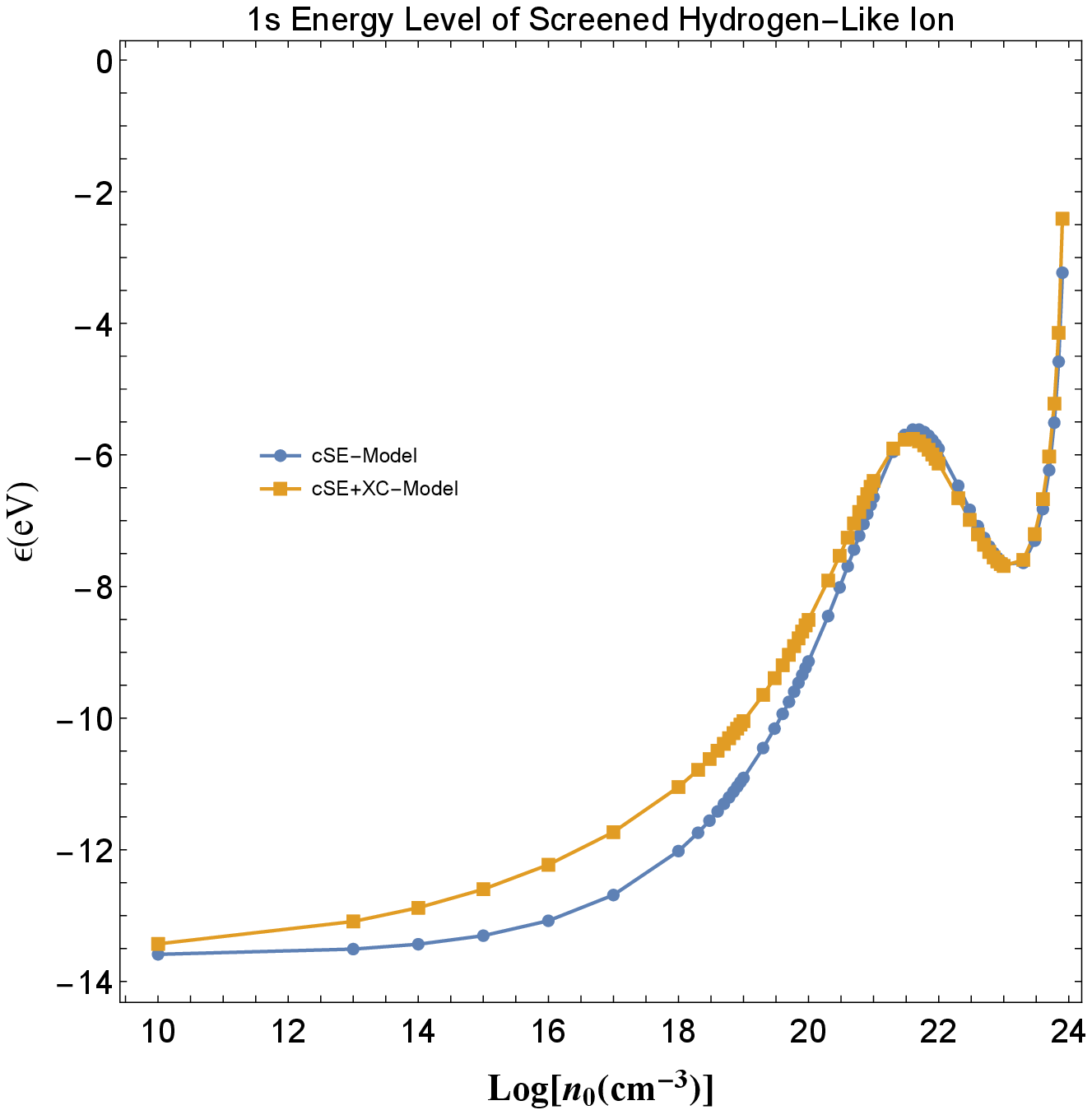}\caption{The variation of 1s energy level of static hydrogen-like ion in a fully degenerate electron gas for different electron number density in a logarithmic scale in the cSE model with and without exchange effect.}
\end{figure}

The asymptotic iteration method (AIM), as a powerful method of solving the second-order differential equations, first has been used by Ciftci et al. \cite{cif1,cif2} in order to obtain the energy eigenvalue problem. The Schr\"{o}dinger equation with different potential profiles has been used with this method to calculate the energy eigenvalues for in a rather simple iterative procedure. The results of calculation by AIM has been confirmed by comparison with other algorithms such as the Ritz variation method. The simple description of the AIM method follows here. Consider the following general second-order homogenous differential equation
\begin{equation}\label{aim}
{y''_n}(r) = {\lambda _0}(r)\,{y'_n}(r) + {s_0}(r)\,y_n(r),
\end{equation}
with ${\lambda _0}(r)\ne 0$ and the functions ${\lambda _0}(r)$ and ${s_0}(r)$ are being sufficiently differentiable with the prime sign denoting the derivative with respect to $r$. The equation (\ref{aim}) has a general solution given as
\begin{equation}\label{sol}
{y_n}(r) = \exp \left( { - \int_{0}^r {\alpha ({r_1})d{r_1}} } \right)\left[ {{C_2} + {C_1}\int_{0}^r {\exp \left( {\int_{0}^{{r_1}} {\left[ {{\lambda _0}({r_2}) + 2\,\alpha ({r_2})} \right]d{r_2}} } \right)d{r_1}} } \right].
\end{equation}
Then, for sufficiently large $n$, the fractional function limit is obtained
\begin{equation}\label{lim}
\frac{{{s_j}(r)}}{{{\lambda _j}(r)}} = \frac{{{s_{j + 1}}(r)}}{{{\lambda _{j + 1}}(r)}} = \alpha (r),
\end{equation}
in which ${\lambda _0}(r)$ and ${s_0}(r)$ follow recursion relations
\begin{subequations}\label{rec}
\begin{align}
&{\lambda _j}(r) = {\lambda '_{j - 1}}(r) + {s_{j - 1}}(r) + {\lambda _0}(r){\lambda _{j - 1}}(r),\\
&{s_j}(r) = {s'_{j - 1}}(r) + {s_0}(r){\lambda _{j - 1}}(r).
\end{align}
\end{subequations}
By definition the following quantization condition leads to the energy eigenvalues
\begin{equation}\label{del}
{\delta _j}(r) = {\lambda _{j + 1}}(r){s_j}(r) - {\lambda _j}(r){s_{j + 1}}(r) = 0,
\end{equation}
where $j$ denotes the iteration number. The radial Schrödinger equation may be transformed into the form (\ref{aim}) for a desired potential function leading to the functional forms ${\lambda _0}(r)$ and ${s_0}(r)$ and the recurrence relations (\ref{rec}) are used to calculate the consequent iterated functions. The energy eigenvalues are then obtained from the roots of (\ref{del}) with the consequent principal quantization numbers $n$ being obtained iteratively from.

The separated radial part of the Schr\"{o}dinger equation may be simplified for eigenfunctions which are related to the wavefunction by $P_{n\ell}(r)=r\Psi_{n\ell}(r)$, as
\begin{equation}\label{se}
\left[ { - \frac{{{\hbar ^2}}}{{2\,m}}\left( {\frac{{{d^2}}}{{d{r^2}}} - \frac{{\ell(\ell  + 1)}}{{{r^2}}}} \right) - V(r)} \right]\,{P_{n\ell }}(r) = {\epsilon_{n\,\ell }}\,{P_{n\ell }}(r),
\end{equation}
in which $n$, $\ell$ are the principal and orbital quantum numbers and $V(r)$ is the ambient electrostatic potential. The transformed wavefunction is taken as the general form of ${P_{n\ell }}(r) = {r^{\ell  + 1}}{e^{ - \kappa r}}f(r)$ in which $f(r)$ denotes an arbitrary function of radial coordinate. The Eq. (\ref{se}) may be normalized in electronvolts units with the radial parameter being normalized to the Bohr radius as follows
\begin{equation}\label{sen}
P''_{n\ell}(r) + \left[ {\frac{\epsilon_{n\ell}}{13.6} - \frac{{l\left( {\ell + 1} \right)}}{{{r^2}}} - 2V(r)} \right]{P_{n\ell}}(r) = 0.
\end{equation}
The characteristic functions $f(r)$ satisfies the following equation
\begin{equation}\label{aim}
{f''}(r) = {\lambda _0}(r)\,{f'}(r) + {s_0}(r)\,f(r),
\end{equation}
with the coefficients given as
\begin{subequations}\label{ls}
\begin{align}
&{\lambda _0}(r) = 2\,\left( {\kappa  - \frac{{\ell + 1}}{r}} \right),\\
&{s_0}(r) = \frac{{\kappa(\ell + 1)}}{r} - {\kappa ^2} - \frac{{{\epsilon_{n\ell}}}}{{13.6}} + 2V(r).
\end{align}
\end{subequations}
The calculation of the energy eigenvalues is possible by the quantization condition, i.e., (\ref{del}) and $V(r)=-\Phi(r)$ defined already for TF, SE and cSE models. While the variables $\epsilon_{n\ell}$ and $r$ both appear in each iteration, by considering the condition $\delta_j(r)=0$ the calculated eigenvalues should be independent from the choice of $r$. The choice of $r$ can be critical for speed of convergence but it also should minimized the potential or equivalently maximized the wave function \cite{soy3}. Therefore, the best choices can be $r=(\ell+1)/\kappa$. In our calculations we used the value of $\kappa=0.6$ for all values of energy with $\ell=0$.

\section{Numerical Analysis and Discussion}

Figure 3 shows the 1s energy level of singly ionized hydrogen-like ion in the electron fluid environment for a wide range of electron number density. Although the used models are based on the complete degeneracy assumption which limits the application of the theory for the classical density region, however, the whole range of density is used for a clear comparison only. The nonrelativistic completely degenerate region starts approximately from $n_0\simeq 10^{20}$cm$^{-3}$ up to $n_0\simeq 10^{24}$cm$^{-3}$. It is evident that the 1s energy level for an isolated hydrogen ion in all models approaches the expected value of $\epsilon_{10}=-13.6$eV. It is also remarked that as the electron density increases initially the level lifts up towards the continuum limit which is also as expected based on the charge screening theory. Note that the increase in the electron number density leads to decrease/increase in the screening length/wavenumber. This consequently gives rise to the weakening of the bond strength of attached electrons and eventually a pressure ionization of neutral atoms take place beyond a critical electron density. It is therefore expected that the hydrogen ion loses its bound energy levels as the electron density is increased. Therefore the monotonic increase in the 1s level for all three models is indeed physical. However, this increase in more rapid for Thomas-Fermi model as compared to two others. The SE and cSE models show remarkable shift of 1s energy level at a critical density which coincides with that for metallic elements and are completely different from the semiclassical Thomas-Fermi model of screening. In the Thomas-Fermi model by increase of the electron density 1s energy level increases monotonically up to the final density of $n_0\simeq 10^{21}$cm$^{-3}$ a point where the 1s level becomes unbound. However, in cSE and SE models the is level increases up to a maximum value beyond which further increase in electron density leads to decrease of 1s level and after reaching a minimum value it lifts again. The values of number density corresponding to these maximum and minimum in cSE model are respectively $n_0\simeq 4.45\times 10^{21}$cm$^{-3}$ and $n_0\simeq 1.28\times 10^{23}$cm$^{-3}$. The 1s level becomes unbound in the cSE model at the density of $n_0\simeq 7.54\times 10^{23}$cm$^{-3}$. Moreover, the values of number density corresponding to the maximum and minimum in SE model are respectively $n_0\simeq 4.1\times 10^{22}$cm$^{-3}$ and $n_0\simeq 6.12\times 10^{24}$cm$^{-3}$. Also, the level becomes completely unbound in the SE model at the number density of $n_0\simeq 2.04\times 10^{25}$cm$^{-3}$. Recent calculation based on the Ritz variational method, implemented for smaller range of electron concentration, does not provide a complete picture for features of SE and cSE potentials and gives rise to totaly different results for 1s energies of hydrogen and helium ions from ours \cite{hu}. The decrease range of 1s energy level of hydrogen-like ion may be attributed to some underlying stabilization mechanism for bound ionic states. Particularly, this stabilizing density range in the cSE model coincides exactly with the metallic bond regime. The stabilization of this type may be due to the interaction of plasmon and Fermi spheres of metallic compounds. However, for SE screening model the density range extends beyond the typical metallic electron density.

Figure 4 shows the 1s energy level corresponding to the bound states of helium-like ion ($Z=2$) for cSE, SE and TF models. It is interesting that for doubly ionized atom the stabilization mechanism which was found for hydrogen-like ions does not take place. This is contrary to findings of the recent literature based on the variational approach \cite{hu}. However, for the SE model a deflection around the electron density of $n_0\simeq 10^{24}$cm$^{-3}$ takes place which leads to a shift of unbound density to higher values compared to the two other models. It is remarked that for isolated in the 1s level has an expected energy of $\epsilon_{10}\simeq -54.4$eV. With increase in the electron density a monotonic increase in 1s level up to a final density is predicted by both TF, SE and cSE models. The density corresponding the points where the helium 1s level becomes completely unbound in TF, SE and cSE models are, respectively, $n_0\simeq 10^{22}$cm$^{-3}$, $n_0\simeq 6.31\times 10^{24}$cm$^{-3}$ and $n_0\simeq 10^{23}$cm$^{-3}$.

In Table. 2 (Fig. 5) we have shown the parameters corresponding to cSE model with and without the electron exchange and correlation effects for a wide range of electron number density. It is remarked that $\alpha$ and $\Gamma$ parameters are significantly affected by these effects and the changes are more pronounced at lower electron density. The variation of the fundamental potential parameter, $\Gamma$, is shown in Fig. 6(a) with respect to the electron number density in cSE model with and without the exchange-correlation effects. It is remarked that this parameter in the presence of exchange-correlation effects crosses twice the critical value $\Gamma=1$ at electron concentrations $n_0\simeq 2.08\times 10^{15}$cm$^{-3}$ and $n_0\simeq 4\times 10^{23}$cm$^{-3}$, respectively. This indicates that the screening potential becomes oscillatory in this electron concentration range. Beyond this range, however, it becomes monotonic. The oscillatory potential is known to lead to oscillatory density variations around the screened impurity charge in metals known as the Frieldel oscillations. In the cSE model the condition $E_p<2E_F$ ($n_0\simeq 1.74\times 10^{23}$cm$^{-3}$) requires the oscillatory shape of screening potential. However, in the generalized model with exchange correlation the condition becomes $E_p>2(E_F+E_{XC})$ with the upper density bound significantly increased to the electron number density of $n_0\simeq 4\times 10^{23}$cm$^{-3}$. Moreover, Fig. 6(b) shows a comparison of the screening parameter $A$ in the cSE model with and without exchange-correlation effects. It is evident that this parameter is greatly effected by these effect. It is remarked that the exchange-correlation effects lead to sharp increase in the screening length over the whole range of electron density but more pronounced in the low electron concentration regime.

Figure 7 depicts the 1s energy level of screened hydrogen-like ion in the cSE model with and without electron exchange-correlation effects. It is evident that exchange-correlation effect alters the energy level variation at lower electron concentration. However, these effect does not alter significantly the stabilization density range. In the presence of the exchange-correlation effects the stabilization takes place in density range from $n_0\simeq 3.69\times 10^{21}$cm$^{-3}$ to $n_0\simeq 1.32\times 10^{23}$cm$^{-3}$ which is slightly lower compared to that without exchange-correlation effects. In the presence of exchange and correlation effects the 1s level becomes unbound in slightly higher electron concentration of $n_0\simeq 8.91\times 10^{23}$cm$^{-3}$. A comparison with real data for concentration of electron in metals \cite{kit} one is convinced that all metallic elements ranging from cesium with the lowest concentration $n_0\simeq 0.9\times 10^{22}$cm$^{-3}$ to beryllium with the highest electron concentration $n_0\simeq 1.2\times 10^{23}$cm$^{-3}$ including transition metals reside in the stable range between the maximum and minimum in Fig. 7. It is interesting to note that metals closest to the minimum is the minimum which is beryllium is the strongest elemental metal with the shortest lattice spacing of $2.22$ angstrom and the furthest one being cesium is the weakest metal with largest lattice spacing of $5.235$ angstrom among all metallic elements. It is remarkable to find that transition metals known to be strong metallic elements have relatively higher electron concentrations with the nickel being the strongest transition metal having the highest electron concentration of $n_0\simeq 9.14\times 10^{22}$cm$^{-3}$ and a lattice parameter $2.49$ angstrom slightly higher than that for beryllium. According to the generalized cSE model the electron concentration corresponding to the turning point of 1s energy level of hydrogen ion at $n_0\simeq 4.45\times 10^{21}$cm$^{-3}$ ($r_s\simeq 7.13$) is where a first-order insulator-to-metal transition takes place.

\section{Conclusion}

We used the asymptotic iteration method to calculate the first energy level of hydrogen and helium-like ions in a degenerate electron gas. Based on three different screening models, namely, the Thomas-Fermi (TF), Shukla-Eliasson (SE) and corrected Shukla-Eliasson (cSE) the variation of the 1s level with the change in the electron number density was compared and remarkable differences and similarities between the first and later two models were coined. The original Thomas-Fermi model predicts a monotonic increase in the 1s level of both hydrogen and helium-like ions as the electron concentration is increased. However, in SE and cSE models of screening the 1s level of hydrogen ion first increases then passes through a maximum and then a minimum and then increases towards the continuum limit. The existence of the corresponding sudden decrease in energy level with increase in the electron concentration which occurs for the exact metallic electron concentration for the cSE model was attributed to a stabilization mechanism which takes place due to the interaction between the plasmon and Fermi energies. In the case of helium-like ions no stabilization of this kind was found. The effect of electron exchange and correlations on variation of 1s energy level of hydrogen-like ion was investigated and has been found that no significant the exchange and correlation have no significant effect on the stable density range. Finally, a simple criterion for metal-insulator transition in the cSE quantum charge screening model is given.

\section{Acknowledgements}

The author A.M.M sincerely appreciates instructive comments of Dr. B. Eliasson.

\section{Data Availability Statement}

The data that support the findings of this study are available from the corresponding author upon reasonable request.

\end{document}